\documentclass[sigconf,nonacm]{acmart}

\usepackage{comment}
\usepackage{booktabs}

\AtBeginDocument{%
  }

\setcopyright{none}
\acmConference[Preprint]{Preprint}{June 2026}{}
\acmBooktitle{Preprint}
\acmDOI{}
\acmISBN{}

\begin{document}

\title{Navigation Alone Is Not Enough: Evaluating Explanatory Assistive UI Agents}

\author{Santosh Patapati}
\email{santosh.patapati@stonybrook.edu}
\affiliation{%
  \institution{Stony Brook University}
  \city{Long Island}
  \state{New York}
  \country{USA}
}

\renewcommand{\shortauthors}{Patapati}

\begin{abstract}
  Modern web interfaces are increasingly difficult to use with screen readers, particularly when pages update dynamically or hide important structure behind visual layout. Recent UI agents can act on such interfaces; however, for assistive agents to be truly useful, they must behave as collaborators that keep users informed and in control, rather than as tools that simply take actions on users’ behalf. Most existing benchmarks judge systems primarily by task completion, without assessing how well they explain their actions or support user oversight. We introduce NeXUI, a benchmark for assistive agents that must navigate interfaces while explaining each step in clear language for nonvisual use. NeXUI pairs realistic user goals with instrumented interface states, enabling agents to reason from both visual context and structural information. Its evaluation measures safety, efficiency, and task success, while also checking whether explanations are grounded in the interface state. In our experiments, we find that NeXUI remains challenging even for state-of-the-art foundation models, with %
  Gemini-3.5-Flash achieving only a 44\% success rate and poor explanation scores, making it a useful foundation for future research and development. By focusing on navigation, explanation, and user control, NeXUI provides a clearer way to study agents that can support blind and visually impaired users in modern computing environments.
\end{abstract}


\maketitle

\section{Introduction}
Modern computing increasingly depends on web interfaces. People use them to manage work, education, health care, finances, and everyday services. Screen readers make many of these tasks possible for blind and visually impaired users. Still, many modern interfaces remain difficult to use in practice. The challenge is often not that the page is completely inaccessible. Instead, users must piece together structure, state, and intent from cues that were designed mainly for visual interaction. This makes routine tasks slower and can make errors harder to notice.

These difficulties are becoming more pronounced as web interfaces become more dynamic. For example, a page may update after a button is pressed without clearly displaying what has changed. In such a case, the user needs more than a list of page elements; they need help understanding what the interface means, what action is available, and what may happen next.

Recent progress in foundation models has made this form of help more plausible. New UI agents can inspect states of an interface and act on natural language goals. This creates an important opportunity for accessible computing; an assistive agent could help a user understand a page update or continue when a workflow becomes unclear. Its value therefore depends on more than its ability to click or type, the user needs to understand what the agent is doing as the task unfolds.

This matters especially for blind and visually impaired users. When an agent acts through an interface, it may change information that the user cannot easily verify. A useful assistant should describe its intent while it works. It should also leave room for the user to approve steps that affect important information or commit an action. This form of collaboration requires the agent to connect the user’s goal with the current interface state, and it must then explain its actions in language that supports oversight.

Benchmarks for UI agents have made this area easier to study as they give models concrete goals and measure whether those goals are reached. This has helped move the field toward more realistic interface tasks. However, task completion alone does not necessarily fully capture what matters in assistive use. Particularly, without clear explanations and collaboration with the user, a system may leave the user unsure about what happened even if it reaches the correct final state.

This creates a need for benchmarks in this area that treat both action and communication as part of the same problem. An assistive agent should be evaluated on how it moves through an interface, how it explains that movement, and how well it preserves user oversight. The benchmark should also reflect the way interfaces are experienced in practice. Agents need to reason from visual appearance as well as interface structure. They also need to recognize when an action requires care. This pushes evaluation more toward the quality of the interaction, rather than only the final outcome.

We introduce NeXUI to help study this directly. NeXUI is a benchmark for assistive agents that must move through an interface while explaining what they are doing in language that is suited to use by low-vision or blind users. Each task in NeXUI combines a realistic user goal with an instrumented interface state. The agent must choose actions that move the task forward, and it must explain those actions in a way the user can follow. This keeps the benchmark focused on the kind of collaboration that assistive use actually requires.

Importantly, NeXUI asks more from a system than simple task completion. It evaluates whether an agent reaches the goal, whether it acts safely, and whether its explanations stay grounded in the current state of the interface. It also captures moments where the right choice is to pause and return control to the user. THis makes the benchmark useful for studying a fuller form of assistive behavior. Particularly, the focus is not just on whether an agent can act, but on whether it can act in a way that users can oversee.

In its current release, NeXUi includes 225 production tasks across 16 interface surfaces. The benchmark draws from accessibility demos, public service prototypes, authenticated web applications, and browser environments. The tasks reflect common kinds of interface work. This includes opening settings, filling forms, recovering form validation errors, and checking whether a change actually took effect. This gives the benchmark realistic interaction settings while keeping evaluation controlled and repeatable.

The contributions of this paper are as follows:

\begin{itemize}
    \item We introduce NeXUI, a benchmark for assistive UI agents that must navigate interfaces while explaining their actions in language suited to nonvisual use.\footnote{Code is available at \url{github.com/Soontosh/NeXUI-Data}}
    \item We present a dataset of 225 tasks across 16 interfaces. The tasks pair realistic user goals with interface states that preserve both visual context and accessibility structure.
    \item We define an evaluation setting that goe sbeyond just task completion. Particularly, NeXUI measures task success, safety, efficiency, and whether an agent's explanations remain grounded in the current interface state.
    \item We study current foundation models on this benchmark and show that NeXUI remains challenging, ensuring that it is useful as a platform for future research.
\end{itemize}

\section{Literature Review}
\subsection{Accessible and Nonvisual Interaction on the Modern Web}
Screen readers provide access to webc ontent by presenting the semantic structure which is exposed by the browser. The process which screen readers follow depends on the existence of meaningful HTML and on accessibility information that communicates the state of interface elements. WAI-ARIA was developed to extend this information for elements that may lack native semantics \cite{w3c2023waiaria}. Examples of this include dynamic content and custom controls. However, accessible markup alone does not guarantee a good experience.  Screen reader interaction often follows a linear reading order, so users rely on headings or search to avoid moving through every element in sequence \cite{borodin2010survey}. When a page does not have clear structure, however, these strategies become much less effective.

Dynamic interfaces create further difficulty here because the page can change while the user's point of attention remains elsewhere. WAI-ARIA provides mechanisms for exposing such changes, but these mechanisms depend on correct implementation, which cannot be reliable across the various implementations of browsers and assistive technologies \cite{w3c2023waiaria}. Recent work on accessibility evaluation has therefore evolved past just checks of the initial page. Demodocus explores reachable states in dynamic web applications so
that accessibility barriers revealed through interaction can be identified \cite{bostic2021demodocus}. This work shows that accessibility must be considered across an interaction rather than only at page load.

Visual layout can also carry meaning that is absent from accessibility structure. Schaadhardt et al. found that blind users working with digital artboards faced high cognitive load and uncertainty about whether object changes had succeeded \cite{schaadhardt2021artboards}. They also had difficulty understanding spatial relationships among objects. Related problems are also seen in online data visualizations. Sharif et al. found that screen reader users spent 211\% more time and were 61\% less accurate than participants who did not use screen readers \cite{sharif2021visualizations}. These results show that access to individual elements does not always provide access to the relationships that make an interface understandable.

Several systems have addressed these problems within specific areas. Azimuth generates dashboard structures alongside descriptions that support screen reader navigation at different levels of detail \cite{srinivasan2023azimuth}. AltCanvas provides a tile-based editor that allows visually impaired
users to create and revise visual scenes with speech and audio feedback \cite{lee2024altcanvas}. Both systems show the value of maintaining some sort of context and giving users direct ways to inspect complex content. However, they are built for specific interface types. %

\subsection{Assistive AI and User Oversight}
Assistive systems are most useful when users can understand and guide their behavior. Prior work on interaction between humans and AI has therefore emphasized clear feedback during use. It has also emphasized the importance of supporting correction when a system makes a mistake \cite{amershi2019guidelines}. These principles are especially relevant for UI agents. In today's web, actions often affect the state of a page before teh user can inspect the result. The agent should therefore communicate enough information for the user to follow its progress and decide whether further action is appropriate.

Recent work has explored how AI can provide support that reflects the user's current interface context. AskEase uses information from the screen and recent interaction history to guide people who use screen readers \cite{chen2026askease}. Its evaluation showed that context aware assistance can improve task completion. Importantly, this is done while reducing the effort required from users. This work supports the broader value of agents that understand both the current interface and the task being performed. It also shows why assistance should be connected to the state of the interaction rather than provided as general advice.

Studies of generative AI use by blind people provide further evidence for the importance of clear and reviewable assistance. Blind and low-vision users apply AI scene-description tools to interpret
visual information in everyday settings \cite{gonzalez2024scene}. They have also developed ways to check uncertain outputs and respond when the system is wrong \cite{alharbi2024misfitting}. These practices show that users remain active participants in the interaction. Assistive agents should therefore provide information that helps users assess their behavior instead of hiding the reasoning behind each step.

\subsection{Web and Computer Use Agent Benchmarks}
Early work on web agents used controlled environments so that interaction could be measured reliably. World of Bits focused on web use as a sequence of keyboard and mouse actions that was guided by natural language\cite{shi2017worldofbits}. MiniWoB++ expanded this approach with tasks that isolate common browser operations\cite{liu2018reinforcement}. WebShop later introduced longer shopping tasks that require an agent to interpret a request and move through several pages \cite{yao2022webshop}. These environments established task success as a practical measure of web agent ability.

More recent benchmarks have moved toward broader and more realistic settings. Mind2Web collects demonstrations from real websites and tests whether agents can transfer across sites and domains \cite{deng2023mind2web}. WebArena provides self-hosted websites with working services and evaluates the state produced by the agent's actions \cite{zhou2023webarena}. WorkArena focuses on tasks performed in enterprise software, where agents carry out routine knowledge work through a browser \cite{drouin2024workarena}. This area of research progress has made evaluation more representative of the workflows found in modern applications.

Other benchmarks have increased the forms of context and interaction available to an agent. VisualWebArena includes tasks where visual information is needed for completion \cite{koh2024visualwebarena}. WebLINX reframes web navigation as a conversation; it records how user instructions develop across several turns \cite{lu2024weblinx}. AssistantBench studies realistic tasks that may require search across the open web \cite{yoran2024assistantbench}.

Together, these benchmarks have made it possible to measure progress on increasingly realistic forms of control for the interface. Their evaluation is usually focused on the result of execution, such as reaching a target state. NeXUI expands on this progress while also studying another requirement. It evaluates navigation together with explanations that need to be grounded in the state of the interface. Therefore, it aims to understand the quality of the interaction alongside successful completion.

\subsection{Multimodal Grounding and Interface Representations}

A UI agent can only act on information that is present in its representation of the interface. Text based representations can contain labels and semantic roles in a form that language models can process directly. They can also reduce the amount of irrelevant page content shown to the model. However, some releationships are expressed through visual placement rather than page structure. Such cases (e.g., changes in color) have led researchers to study both structural and visual views of an interface.

Several web agent frameworks rely on structured page information to support the selection of actions. Mind2Web, for example, grounds actions to elements taken from the underlying webpage \cite{deng2023mind2web}. WebArena also uses structured observations so that agents can refer to page elements during execution \cite{zhou2023webarena}. BrowserGym extends this approach by supporting screenshots alongside HTML and the browser accessibility tree \cite{drouin2024workarena}. These representations provide semantic information that may be difficult to recover from pixels alone. They also allow actions to target stable element references instead of relying only on screen coordinates.

Visual models provide another view of the interface. ScreenAI was trained to unerstand the layout and content of user interfaces from screenshots \cite{baechler2024screenai}. VisualWebArena showed that some web tasks cannot be solved reliably without attending to visual information \cite{koh2024visualwebarena}. SeeAct further studied how multimodal models connect language instructions to elements on live websites \cite{zheng2024seeact}. Its results show that high-level reasoning can be strong even when the correct element is difficult to locate. The most effective grounding method in that study used both visual information and page structure, which suggests that the two forms of context can support one another.

NeXUI follows this broader direction while using mutlimodal interface states for a goal focused on accessibility. Its tasks preserve the rendered page together with structural information from the browser. They also provide a reader oriented view of the content and a set of available action targets. These views allow an agent to consider how the page appears while reasoning about the information available. This information is exposed in such a way that is accessible to visually impaired people. NeXUI then evaluates whether the selected action and its explanation are grounded in that shared state. In this way, interface representation supports both navigation and communication with the user.

\subsection{Evaluation Beyond Task Completion}
Most UI agent benchmarks use task completion as their main measure of performance. WebArena checks whether the agent reaches the required state after interacting with a website \cite{zhou2023webarena}. VisualWebArena applies a similar approach to tasks that depend on visual information \cite{koh2024visualwebarena}. AssistantBench also uses automatically checked outcomes for longer tasks on the open web \cite{yoran2024assistantbench}. These measures provide a clear way to compare systems; they have supported steady progress in web agent research.

Other benchmarks inspect parts of the interaction before the final outcome. Mind2Web evaluates whether a model selects the correct element and action at each step \cite{deng2023mind2web}. WebLINX studies action prediction within conversations between a user and a web agent \cite{lu2024weblinx}.

\section{The NeXUI Benchmark}
\subsection{Navigate and Explain Tasks}
Each NeXUI task begins with a user gaol and a captured state of an interface. The goal describes what the user wants to accomplish in everyday language. At each step, the agent receives information about the current page and the actions that are available. It then chooses an action that moves the task forward. The resulting state is returned to the agent so that it can continue until the goal has been reached.

NeXUI also requires the agent to explain every action it takes. The explanation should tell the user what the agent is doing and how the step relates to the task. It must remain grounded in the current interface state. In particular, the agent should not claim that a change has occured before that change can be observed. This requirement allows the benchmark to study whether an agent can make progress while keeping its behavior clear to a visually impaired person that is following the task through nonvisual interaction.

Some tasks contain actions that should not be completed without approval. In these cases, the correct response is to pause and ask the user whether the agent should continue. For example, one task asks the agent to prepare a payment for a contact. The agent enters the payment details but must stop before pressing the final paymet button; pausing at this point counts as successful behavior because it keeps the user's control over the interaction. Tasks without such a boundary end when the agent reaches the necessary state and provides a final summary.

\subsection{Dataset Coverage}
The current release of NeXUI contains 225 production tasks across 16 interfaces. These tasks include 1,755 captured interface states and more than 31,000 candidate targets. Most tasks unfold over several steps, which allows the benchmark to test whether an agent can maintain the user's goal as the interface changes. The dataset also contains 43 tasks with a confirmation boundary. In these cases, successful behavior requires the agent to stop before carrying out an important final action.

The tasks are drawn from several kinds of web environments. Some come from accessibility demonstrations that expose common barriers in page structure. Others use public service prototypes or authenticated applications. The benchmark also includes enterprise workflows. This range enables more robust evaluation on teh dataset. It covers interfaces that differ in layout and style of interaction.

Table~\ref{tab:task-counts} summarizes the split sizes of NeXUI.

\begin{table}[t]
\centering
\caption{Task counts in the NeXUI splits. The challenge set is a subset of the test split.}
\label{tab:task-counts}
\begin{tabular}{lc}
\toprule
Split & Number of tasks \\
\midrule
Development & 146 \\
Validation & 25 \\
Test & 54 \\
Challenge & 14 \\
\bottomrule
\end{tabular}
\end{table}

Tasks are also labeled by difficulty. The current release contains 23 easy tasks and 29 medium tasks. It also includes 128 hard tasks and 45 very hard tasks. These labels support analysis of how model performance changes as tasks require longer interaction or more careful reasoning.

\subsection{Interface States and Task Annotations}

Each task contains a sequence of captured interface states. Every state preserves the rendered page as well as information drawn from the browser structure. NeXUI also records the accessibility tree and a reader oriented view of the page. These representations provide different forms of context. The rendered page retains visual layout, while the structural views expose labels and relationships that may be important for nonvisual interaction

The benchmark also identifies the actions available in each state. Each candidate includes a stable reference and information about the corresponding interface element. When the agent submits an action, NeXUI checks it against the transition encoded for the current state. A supported action advances the task to the next captured state. Actions outside the encoded transition do not create an alternative path.

Each task includes a reference trajectory that defines the supported path toward the goal. The packaged tasks use linear transition graphs, so the agent must follow the encoded sequence of actions and states. The trajectory also records an explanatoin for each step. Conditions tied to the final state are then used to determine whether the task was completed successfully.

NeXUI provides further annotations for safety and explanation quality. Safety rules identify actions that are forbidden or require approval from the user. Explanation rubrics describe what the agent should communicate during the task. These annotations make it posible to evaluate the full interaction rather than only its final state.

\subsection{Evaluation}

NeXUI evaluates each run across task success, safety, efficiency, and explanation quality. A run is counted as passed only when the task is completed without a critical safety violation. These measures are reported separately so that a model's final outcome can be considered alongside the interaction that produced it.

Task success is determined through conditions tied to the requested goal. Depending on the task, these conditions may check the current page or the value of a field. They may also verify that the agent stopped at the correct point. This allows NeXUI to evaluate the intended outcome.

Safety is evaluated through rules defined for each task. Some rules identify actions that should never be taken. Others mark actions that require confirmation from the user. If the agent performs one of these actions without first returning control, the run receives a critical safety violation. Coordinate-based clicks are also recorded because they provide less reliable grounding than actions tied to a known interface element.

Efficiency is measured by comparing the number of agent steps with the length of the reference trajectory. An agent receives full efficiency credit when it completes the task in no more steps than the reference. Longer runs receive a lower score. The score is reduced further when the task is not completed, since a short unsuccessful run should not be treated as efficient behavior.

Explanation quality is also measured at each step. The scorer considers whether the explanation aligns with the selected action, refers to information int he current state, and connects the action to the user's goal. When structured justification is provided, NeXUI can compare claims about hte expected result with the next captured state. It also measures whether the agent communicates relevant confirmation boundaries and keeps its explanation concise. The strength of these checks depends on the explanation format and the rubric used for the task.

\section{Experimental Evaluation}

\subsection{Models and Experimental Setup}

We evaluate three foundation models from the Gemini Flash family. These are Gemini 2.5 Flash Lite, Gemini 2.5 Flash, and Gemini 3.5 Flash. These models provide a range of cost and reasoning capacity while sharing the same general interface. Our goal is to measure how well current models can complete NeXUI tasks while producing explanations that remain grounded in the interaction.

The experiments use the 25 tasks in the validation split. Each model is evaluated through the text based baseline provided with NeXUI. The baseline does not supply the page screenshot to the model. Instead, the model receieves the user goal and information about the current interface state. This includes the available action targets and recent interaction history. The stronger input setting also provides the reader view and the ARIA representation of the current page.

At each step, the model returns a structured action and a short explanation. The step limit scales with the length of the reference trajectory and is capped at 50 steps. A run ends when the model finishes the task, requests user input, submits an invalid action, or reaches the step limit.

\subsection{Main Results}

Table~\ref{tab:main_results} reports the results on the 25-task validation split. All three models completed every scheduled run. Gemini 3.5 Flash achieved the strongest overall performance, with a task success rate of 44.0\% and a mean explanation score of 0.7359. Gemini 2.5 Flash followed closely at 40.0%

\begin{table}[t]
\centering
\caption{Results on the 25-task NeXUI validation split.}
\label{tab:main_results}
\begin{tabular}{lccc}
\toprule
Model & Runs completed & Success & Explanation \\
\midrule
Gemini 2.5 Flash Lite & 25/25 & 4.0\% & 0.4008 \\
Gemini 2.5 Flash & 25/25 & 40.0\% & 0.6770 \\
Gemini 3.5 Flash & 25/25 & 44.0\% & 0.7359 \\
\bottomrule
\end{tabular}
\end{table}

Explanation performance followed the same overall ordering as task success. Gemini 3.5 Flash received the highest explanation score, followed by Gemini 2.5 Flash. However, neither model solved a majority of the tasks. The results therefore show that stronger explanations do not remove the broader difficulty of reliable task completion. They also demonstrate that NeXUI remains challenging even for recent foundation models.

\section{Discussion and Conclusion}
Our results show that current foundation models can complete some NeXUI tasks, but reliable assistive behavior remains difficult. The strongest model completed 44\% of the validation tasks and also received low explanation scores. This suggests that progress in general UI interaction does not yet translate directly intonts that can communicate their behavior clearly. NeXUI makes this gap easier to study by evaluating navigation and explanation within the same interaction.

Future work can extend NeXUI with more interfaces and richer transition graphs. It can also study agents that use screenshtos together with accessibility structure. These directions may help researchers develop systems that track interface changes more reliably while also communicating more clearly. NeXUI provides an initial foundation for this work by placing task completion, explanations, and user control within a single benchmark.

\bibliographystyle{ACM-Reference-Format}
\bibliography{references}

@techreport{w3c2023waiaria,
  title       = {Accessible Rich Internet Applications ({WAI-ARIA}) 1.2},
  author      = {Diggs, Joanmarie and Nurthen, James and Cooper, Michael and MacLeod, Carolyn},
  institution = {World Wide Web Consortium},
  type        = {{W3C} Recommendation},
  year        = {2023},
  month       = jun,
  url         = {https://www.w3.org/TR/wai-aria-1.2/},
  note        = {Published 6 June 2023}
}

@inproceedings{borodin2010survey,
  title     = {A Survey of Screen-Reader Browsing Strategies},
  author    = {Borodin, Yevgen and Bigham, Jeffrey P. and Dausch, Glenn and Ramakrishnan, I. V.},
  booktitle = {Proceedings of the 7th International Cross-Disciplinary Conference on Web Accessibility},
  series    = {W4A '10},
  year      = {2010},
  publisher = {Association for Computing Machinery},
  address   = {New York, NY, USA},
  doi       = {10.1145/1805986.1806005},
  url       = {https://doi.org/10.1145/1805986.1806005}
}

@misc{bostic2021demodocus,
  title         = {Automated Evaluation of Web Site Accessibility Using a Dynamic Accessibility Measurement Crawler},
  author        = {Bostic, Trevor and Stanley, Jeffrey and Higgins, John and Chudnov, Daniel and Brunelle, Justin and Tracy, Brittany},
  year          = {2021},
  eprint        = {2110.14097},
  archivePrefix = {arXiv},
  primaryClass  = {cs.CY},
  doi           = {10.48550/arXiv.2110.14097},
  url           = {https://arxiv.org/abs/2110.14097}
}

@inproceedings{schaadhardt2021artboards,
  title     = {Understanding Blind Screen-Reader Users' Experiences of Digital Artboards},
  author    = {Schaadhardt, Anastasia and Hiniker, Alexis and Wobbrock, Jacob O.},
  booktitle = {Proceedings of the 2021 CHI Conference on Human Factors in Computing Systems},
  series    = {CHI '21},
  year      = {2021},
  publisher = {Association for Computing Machinery},
  address   = {New York, NY, USA},
  doi       = {10.1145/3411764.3445442},
  url       = {https://doi.org/10.1145/3411764.3445442}
}

@inproceedings{sharif2021visualizations,
  title     = {Understanding Screen-Reader Users' Experiences with Online Data Visualizations},
  author    = {Sharif, Ather and Chintalapati, Sanjana Shivani and Wobbrock, Jacob O. and Reinecke, Katharina},
  booktitle = {Proceedings of the 23rd International ACM SIGACCESS Conference on Computers and Accessibility},
  series    = {ASSETS '21},
  year      = {2021},
  publisher = {Association for Computing Machinery},
  address   = {New York, NY, USA},
  doi       = {10.1145/3441852.3471202},
  url       = {https://doi.org/10.1145/3441852.3471202}
}

@inproceedings{srinivasan2023azimuth,
  title     = {Azimuth: Designing Accessible Dashboards for Screen Reader Users},
  author    = {Srinivasan, Arjun and Harshbarger, Tim and Hilliker, Darrell and Mankoff, Jennifer},
  booktitle = {Proceedings of the 25th International ACM SIGACCESS Conference on Computers and Accessibility},
  series    = {ASSETS '23},
  year      = {2023},
  publisher = {Association for Computing Machinery},
  address   = {New York, NY, USA},
  numpages  = {16},
  doi       = {10.1145/3597638.3608405},
  url       = {https://doi.org/10.1145/3597638.3608405}
}

@misc{lee2024altcanvas,
  title         = {AltCanvas: A Tile-Based Image Editor with Generative {AI} for Blind or Visually Impaired People},
  author        = {Lee, Seonghee and Kohga, Maho and Landau, Steve and O'Modhrain, S{\'i}le and Subramonyam, Hari},
  year          = {2024},
  eprint        = {2408.10240},
  archivePrefix = {arXiv},
  primaryClass  = {cs.HC},
  doi           = {10.48550/arXiv.2408.10240},
  url           = {https://arxiv.org/abs/2408.10240}
}

@inproceedings{amershi2019guidelines,
  title     = {Guidelines for Human-{AI} Interaction},
  author    = {Amershi, Saleema and Weld, Dan and Vorvoreanu, Mihaela and Fourney, Adam and Nushi, Besmira and Collisson, Penny and Suh, Jina and Iqbal, Shamsi T. and Bennett, Paul N. and Inkpen, Kori and Teevan, Jaime and Kikin-Gil, Ruth and Horvitz, Eric},
  booktitle = {Proceedings of the 2019 CHI Conference on Human Factors in Computing Systems},
  series    = {CHI '19},
  year      = {2019},
  publisher = {Association for Computing Machinery},
  address   = {New York, NY, USA},
  doi       = {10.1145/3290605.3300233},
  url       = {https://doi.org/10.1145/3290605.3300233}
}

@misc{chen2026askease,
  title         = {From Struggle to Success: Context-Aware Guidance for Screen Reader Users in Computer Use},
  author        = {Chen, Nan and Lu, Jing and Wang, Zilong and Qiu, Luna K. and Chen, Siming and Yang, Yuqing},
  year          = {2026},
  eprint        = {2601.18092},
  archivePrefix = {arXiv},
  primaryClass  = {cs.HC},
  doi           = {10.48550/arXiv.2601.18092},
  url           = {https://arxiv.org/abs/2601.18092}
}

@misc{gonzalez2024scene,
  title         = {Investigating Use Cases of {AI}-Powered Scene Description Applications for Blind and Low Vision People},
  author        = {Gonzalez, Ricardo and Collins, Jazmin and Azenkot, Shiri and Bennett, Cynthia},
  year          = {2024},
  eprint        = {2403.15604},
  archivePrefix = {arXiv},
  primaryClass  = {cs.HC},
  doi           = {10.48550/arXiv.2403.15604},
  url           = {https://arxiv.org/abs/2403.15604}
}

@misc{alharbi2024misfitting,
  title         = {Misfitting With {AI}: How Blind People Verify and Contest {AI} Errors},
  author        = {Alharbi, Rahaf and Lor, Pa and Herskovitz, Jaylin and Schoenebeck, Sarita and Brewer, Robin},
  year          = {2024},
  eprint        = {2408.06546},
  archivePrefix = {arXiv},
  primaryClass  = {cs.HC},
  doi           = {10.48550/arXiv.2408.06546},
  url           = {https://arxiv.org/abs/2408.06546}
}

@inproceedings{shi2017worldofbits,
  title     = {World of Bits: An Open-Domain Platform for Web-Based Agents},
  author    = {Shi, Tianlin and Karpathy, Andrej and Fan, Linxi and Hernandez, Jonathan and Liang, Percy},
  booktitle = {Proceedings of the 34th International Conference on Machine Learning},
  series    = {Proceedings of Machine Learning Research},
  volume    = {70},
  pages     = {3135--3144},
  year      = {2017},
  publisher = {PMLR},
  url       = {https://proceedings.mlr.press/v70/shi17a.html}
}

@inproceedings{liu2018reinforcement,
  title     = {Reinforcement Learning on Web Interfaces Using Workflow-Guided Exploration},
  author    = {Liu, Evan Zheran and Guu, Kelvin and Pasupat, Panupong and Shi, Tianlin and Liang, Percy},
  booktitle = {International Conference on Learning Representations},
  year      = {2018},
  eprint    = {1802.08802},
  archivePrefix = {arXiv},
  url       = {https://arxiv.org/abs/1802.08802}
}

@inproceedings{yao2022webshop,
  title     = {WebShop: Towards Scalable Real-World Web Interaction with Grounded Language Agents},
  author    = {Yao, Shunyu and Chen, Howard and Yang, John and Narasimhan, Karthik},
  booktitle = {Advances in Neural Information Processing Systems},
  volume    = {35},
  year      = {2022},
  eprint    = {2207.01206},
  archivePrefix = {arXiv},
  url       = {https://arxiv.org/abs/2207.01206}
}

@inproceedings{deng2023mind2web,
  title     = {Mind2Web: Towards a Generalist Agent for the Web},
  author    = {Deng, Xiang and Gu, Yu and Zheng, Boyuan and Chen, Shijie and Stevens, Samuel and Wang, Boshi and Sun, Huan and Su, Yu},
  booktitle = {Advances in Neural Information Processing Systems},
  volume    = {36},
  year      = {2023},
  eprint    = {2306.06070},
  archivePrefix = {arXiv},
  url       = {https://arxiv.org/abs/2306.06070}
}

@misc{zhou2023webarena,
  title         = {WebArena: A Realistic Web Environment for Building Autonomous Agents},
  author        = {Zhou, Shuyan and Xu, Frank F. and Zhu, Hao and Zhou, Xuhui and Lo, Robert and Sridhar, Abishek and Cheng, Xianyi and Ou, Tianyue and Bisk, Yonatan and Fried, Daniel and Alon, Uri and Neubig, Graham},
  year          = {2023},
  eprint        = {2307.13854},
  archivePrefix = {arXiv},
  primaryClass  = {cs.AI},
  doi           = {10.48550/arXiv.2307.13854},
  url           = {https://arxiv.org/abs/2307.13854}
}

@misc{drouin2024workarena,
  title         = {WorkArena: How Capable Are Web Agents at Solving Common Knowledge Work Tasks?},
  author        = {Drouin, Alexandre and Gasse, Maxime and Caccia, Massimo and Laradji, Issam H. and Del Verme, Manuel and Marty, Tom and Boisvert, L{\'e}o and Thakkar, Megh and Cappart, Quentin and Vazquez, David and Chapados, Nicolas and Lacoste, Alexandre},
  year          = {2024},
  eprint        = {2403.07718},
  archivePrefix = {arXiv},
  primaryClass  = {cs.AI},
  doi           = {10.48550/arXiv.2403.07718},
  url           = {https://arxiv.org/abs/2403.07718}
}

@misc{koh2024visualwebarena,
  title         = {VisualWebArena: Evaluating Multimodal Agents on Realistic Visual Web Tasks},
  author        = {Koh, Jing Yu and Lo, Robert and Jang, Lawrence and Duvvur, Vikram and Lim, Ming Chong and Huang, Po-Yu and Neubig, Graham and Zhou, Shuyan and Salakhutdinov, Ruslan and Fried, Daniel},
  year          = {2024},
  eprint        = {2401.13649},
  archivePrefix = {arXiv},
  primaryClass  = {cs.AI},
  doi           = {10.48550/arXiv.2401.13649},
  url           = {https://arxiv.org/abs/2401.13649}
}

@misc{lu2024weblinx,
  title         = {WebLINX: Real-World Website Navigation with Multi-Turn Dialogue},
  author        = {L{\`u}, Xing Han and Kasner, Zden{\v{e}}k and Reddy, Siva},
  year          = {2024},
  eprint        = {2402.05930},
  archivePrefix = {arXiv},
  primaryClass  = {cs.CL},
  doi           = {10.48550/arXiv.2402.05930},
  url           = {https://arxiv.org/abs/2402.05930}
}

@misc{yoran2024assistantbench,
  title         = {AssistantBench: Can Web Agents Solve Realistic and Time-Consuming Tasks?},
  author        = {Yoran, Ori and Amouyal, Samuel Joseph and Malaviya, Chaitanya and Bogin, Ben and Press, Ofir and Berant, Jonathan},
  year          = {2024},
  eprint        = {2407.15711},
  archivePrefix = {arXiv},
  primaryClass  = {cs.CL},
  doi           = {10.48550/arXiv.2407.15711},
  url           = {https://arxiv.org/abs/2407.15711}
}

@misc{baechler2024screenai,
  title         = {ScreenAI: A Vision-Language Model for {UI} and Infographics Understanding},
  author        = {Baechler, Gilles and Sunkara, Srinivas and Wang, Maria and Zubach, Fedir and Mansoor, Hassan and Etter, Vincent and C{\u{a}}rbune, Victor and Lin, Jason and Chen, Jindong and Sharma, Abhanshu},
  year          = {2024},
  eprint        = {2402.04615},
  archivePrefix = {arXiv},
  primaryClass  = {cs.CV},
  doi           = {10.48550/arXiv.2402.04615},
  url           = {https://arxiv.org/abs/2402.04615}
}

@misc{zheng2024seeact,
  title         = {{GPT-4V}(ision) Is a Generalist Web Agent, if Grounded},
  author        = {Zheng, Boyuan and Gou, Boyu and Kil, Jihyung and Sun, Huan and Su, Yu},
  year          = {2024},
  eprint        = {2401.01614},
  archivePrefix = {arXiv},
  primaryClass  = {cs.AI},
  doi           = {10.48550/arXiv.2401.01614},
  url           = {https://arxiv.org/abs/2401.01614}
}

\end{document}